# Exact Solution for Partition function of General Ising Model in Magnetic Fields and Bayesian Networks

Akira Saito 649-202 Kidera, Nara, Nara 630-8306, Japan: tsunagari LLC (Japan)

(http://tsunagari-net.info/)

We propose a method for generalizing the Ising model in magnetic fields and calculating the partition function (exact solution) for the Ising model of an arbitrary shape. Specifically, the partition function is calculated using matrices that are created automatically based on the structure of the system. By generalizing this method, it becomes possible to calculate the partition function of various crystal systems (network shapes) in magnetic fields when N (scale) is infinite. Furthermore, we also connect this method for finding the solution to the Ising model in magnetic fields to a method for finding the solution to Bayesian networks in information statistical mechanics (applied to data mining, machine learning, and combinatorial optimization).

## 1. Introduction

It is believed that the ability to obtain the partition function for the Ising model of an arbitrary shape (from 2-dimensional lattices to general crystal systems and arbitrary network systems) in a magnetic field using only minimal calculations would be a great contribution to the field of information science, due to the strong connection of this problem with solid-state physics and Bayesian networks (involved in several areas that have recently attracted widespread attention, such as data mining, machine learning, and combinatorial optimization). Below, we describe a method for finding the solution of the partition function for a general Ising model in a magnetic field, and apply this method to Bayesian networks.

## 2. Theory

*2.1 Ising model in external magnetic field*

The partition functions for Ising models in the field of statistical mechanics for 1-dimensional and 2-dimensional lattices are well known. I pursued a solution for deriving the partition function of general crystal shapes and arbitrary shapes in a magnetic field for the Ising model.

First, let edges be created between nodes if there are interactions between the nodes. If the nodes at the ends of the edge are (+1,+1) or (-1,-1), then the energy is represented as -Jij. If the nodes at the ends of the edge are (+1,-1) or (-1,+1), then the energy is represented as +Jij. Furthermore, if the external magnetic field is +1, then the energy is represented as -Ji, and if the external magnetic field is -1, then the energy is represented as +Ji. Then, the Hamiltonian can be represented by the following equation:

$$H = \sum_i -J_i \sigma_i + \sum_{\{i,j\}} -J_{i,j}\sigma_i\sigma_j. \qquad (1)$$

The sum over {ij} is taken over all edges. The partition function of the system can be written as follows:

$$Z = \sum_{\{\sigma\}} \prod_i \exp\left(\frac{\sigma_i J_i}{k_B T}\right) \prod_{\{i,j\}} \exp\left(\frac{\sigma_i \sigma_j J_{i,j}}{k_B T}\right). \qquad (2)$$

The sum over {σ} is taken over all values of ±1 of all σ. Further, the equation can be written as follows:

$$Z = \sum_{\{\sigma\}} \prod_i \Big[\tfrac{1}{2}\big\{\exp\left(\tfrac{J_i}{k_B T}\right) + \exp\left(\tfrac{-J_i}{k_B T}\right)\big\}$$

$$+ \tfrac{1}{2}\sigma_i\big\{\exp\left(\tfrac{J_i}{k_B T}\right) - \exp\left(\tfrac{-J_i}{k_B T}\right)\big\}\Big] \prod_{\{i,j\}} \Big[\tfrac{1}{2}\big\{\exp\left(\tfrac{J_{i,j}}{k_B T}\right) + \exp\left(\tfrac{-J_{i,j}}{k_B T}\right)\big\}$$

$$+ \tfrac{1}{2}\sigma_i\sigma_j\big\{\exp\left(\tfrac{J_{i,j}}{k_B T}\right) - \exp\left(\tfrac{-J_{i,j}}{k_B T}\right)\big\}\Big]. \qquad (3)$$

This can be again written as follows:

$$Z = \sum_{\{\sigma\}} \prod_i [\psi_i + \sigma_i\hat{\psi}_i] \prod_{\{i,j\}} [\phi_{i,j} + \sigma_i\sigma_j\hat{\phi}_{i,j}]. \qquad (4)$$

$$\psi_i = \frac{1}{2}\left\{\exp\left(\frac{J_i}{k_BT}\right) + \exp\left(\frac{-J_i}{k_BT}\right)\right\}, \quad \hat{\psi}_i = \frac{1}{2}\left\{\exp\left(\frac{J_i}{k_BT}\right) - \exp\left(\frac{-J_i}{k_BT}\right)\right\}. \quad (5)$$

$$\phi_{i,j} = \frac{1}{2}\left\{\exp\left(\frac{J_{i,j}}{k_BT}\right) + \exp\left(\frac{-J_{i,j}}{k_BT}\right)\right\}, \quad \hat{\phi}_{i,j} = \frac{1}{2}\left\{\exp\left(\frac{J_{i,j}}{k_BT}\right) - \exp\left(\frac{-J_{i,j}}{k_BT}\right)\right\}. \quad (6)$$

Furthermore,

$$Z = \prod_i [\psi_i] \prod_{\{i,j\}} [\phi_{i,j}] \sum_{\{\sigma\}} \prod_i [1 + \sigma_i B_i] \prod_{\{i,j\}} [1 + \sigma_i \sigma_j A_{i,j}]. \quad (7)$$

$$B_i = \hat{\psi}_i / \psi_i. \quad (8)$$

$$A_{i,j} = \hat{\phi}_{i,j} / \phi_{i,j}. \quad (9)$$

Examining Eq. (7), it can be seen that σ takes a value of either +1 or -1, and that this equation takes the sum over all combinations of σ. If the product of i and {ij} contains σ left over, then that term is canceled out as the sum of +1 and -1. Therefore, the only terms that remain are the terms without σ, such as terms which contain the square of σ, which is 1. Each appearance of σ is part of a pair, so terms in which σ does not remain are terms that are combinations in which each number that appears in ij, the subscript of A, and i, the subscript of B, always appears an even number of times for each term in the product of A and B. From Eq. (7), the factor (1+B) exists for each node. Therefore, no matter what combination is used in the product (1+A), it is possible to remove σ from the term by appropriately choosing the value of B. In other words, in terms that include (1+A) as a factor, it is sufficient to consider how it is being multiplied with B (in order to ensure that σ does not remain).

*2.2 Calculation method using matrices*

Next, we will describe the calculation method for the following system.

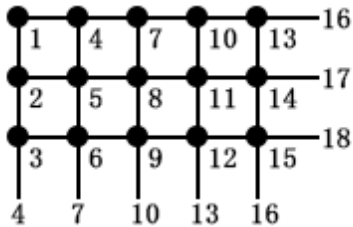

(10)

This is a system in which there are 3×N nodes. Each node has edges extending to the node that is one number ahead and three numbers ahead, so that the system forms a spiral. The numbering is done as shown in Eq. (10). Then, we will consider multiplication by (1+A) in the +1 direction and the +3 direction in the order of the numbering of the nodes.

$$(1+A_{n,n+1})(1+A_{n,n+N}) \tag{11}$$

We multiply from equation (11) for n in order, starting with 1. Once we have multiplied by Eq. (11) for n=n, there is no longer any A remaining that contains n in the subscript. Therefore, if the number of times that n is in the subscript of A in each term is even, then we should multiply by 1, and if it is odd, then we should multiply by Bn.

$$\times 1 \quad \text{for Even} \{A_{n,n-N}, A_{n,n-1}, A_{n,n+1}, A_{n,n+N}\}$$
$$\times B_n \quad \text{for Odd} \{A_{n,n-N}, A_{n,n-1}, A_{n,n+1}, A_{n,n+N}\} \tag{12}$$

The combinations of numbers which are still undetermined before and after multiplying by Eq. (11) are shown as follows:

$$\{\},\{n\},\{n+1\},\{n,n+1\},\{n+2\},\{n,n+2\},\{n+1,n+2\},\{n,n+1,n+2\}$$
$$\Downarrow$$
$$\{\},\{n+1\},\{n+2\},\{n+1,n+2\},\{n+3\},\{n+1,n+3\},\{n+2,n+3\},\{n+1,n+2,n+3\}$$
$$\tag{13}$$

We create a matrix of the product of Eqs. (11) and (12) with its rows and columns given by Eq. (13). By multiplying the factors from n = 1 to n = 3N, we derive the partition function in Eq. (10). This matrix is represented as follows:

$$\begin{pmatrix} & \{\},\{n\},\{n+1\},\{n,n+1\},\{n+2\},\{n,n+2\},\{n+1,n+2\},\{n,n+1,n+2\} \\ \{\} & X_{11}\ X_{12}\ X_{13}\ X_{14}\ X_{15}\ X_{16}\ X_{17}\ X_{18} \\ \{n+1\} & X_{21}\ X_{22}\ X_{23}\ X_{24}\ X_{25}\ X_{26}\ X_{27}\ X_{28} \\ \{n+2\} & X_{31}\ X_{32}\ X_{33}\ X_{34}\ X_{35}\ X_{36}\ X_{37}\ X_{38} \\ \{n+1,n+2\} & X_{41}\ X_{42}\ X_{43}\ X_{44}\ X_{45}\ X_{46}\ X_{47}\ X_{48} \\ \{n+3\} & X_{51}\ X_{52}\ X_{53}\ X_{54}\ X_{55}\ X_{56}\ X_{57}\ X_{58} \\ \{n+1,n+3\} & X_{61}\ X_{62}\ X_{63}\ X_{64}\ X_{65}\ X_{66}\ X_{67}\ X_{68} \\ \{n+2,n+3\} & X_{71}\ X_{72}\ X_{73}\ X_{74}\ X_{75}\ X_{76}\ X_{77}\ X_{78} \\ \{n+1,n+2,n+3\} & X_{81}\ X_{82}\ X_{83}\ X_{84}\ X_{85}\ X_{86}\ X_{87}\ X_{88} \end{pmatrix} = X$$

(14)

$$X_{11}=X_{15}=X_{23}=X_{27}=X_{31}=X_{35}=X_{43}=X_{47} = 1$$
$$X_{12}=X_{16}=X_{24}=X_{28}=X_{32}=X_{36}=X_{44}=X_{48} = B_n$$
$$X_{14}=X_{18}=X_{22}=X_{26}=X_{34}=X_{38}=X_{42}=X_{46} = A_{nn+1}$$
$$X_{13}=X_{17}=X_{21}=X_{25}=X_{33}=X_{37}=X_{41}=X_{45} = A_{nn+1}B_n$$
$$X_{52}=X_{56}=X_{64}=X_{68}=X_{72}=X_{76}=X_{84}=X_{88} = A_{nn+3}$$
$$X_{51}=X_{55}=X_{63}=X_{67}=X_{71}=X_{75}=X_{83}=X_{87} = A_{nn+3}B_n$$
$$X_{53}=X_{57}=X_{61}=X_{65}=X_{73}=X_{77}=X_{81}=X_{85} = A_{nn+3}A_{nn+1}$$
$$X_{54}=X_{58}=X_{62}=X_{66}=X_{74}=X_{78}=X_{82}=X_{86} = A_{nn+3}A_{nn+1}B_n$$

(15)

This matrix is made up of a combination of the following matrices.

$$A_{nn+3} + A_{nn+1} + n = B_n \quad \text{Even} \ \square \quad \text{Odd} \ \blacksquare$$

(16)

Matrix A can be represented as the difference of the following scales.

$$A_{nn+3} \quad A_{nn+2} \quad A_{nn+1}$$

(17)

Matrix B is created by adding the corresponding components of A and n to the connected matrix. If the result is even, then the value becomes 1, and if it is odd, then the value becomes B. If A and B do not depend on ij, then Eq. (15) can be written as

$$\begin{aligned}
X_{11}=X_{15}=X_{23}=X_{27}=X_{31}=X_{35}=X_{43}=X_{47} &= 1 \\
X_{12}=X_{16}=X_{24}=X_{28}=X_{32}=X_{36}=X_{44}=X_{48} &= B \\
X_{14}=X_{18}=X_{22}=X_{26}=X_{34}=X_{38}=X_{42}=X_{46} &= A \\
X_{13}=X_{17}=X_{21}=X_{25}=X_{33}=X_{37}=X_{41}=X_{45} &= AB \\
X_{52}=X_{56}=X_{64}=X_{68}=X_{72}=X_{76}=X_{84}=X_{88} &= A \\
X_{51}=X_{55}=X_{63}=X_{67}=X_{71}=X_{75}=X_{83}=X_{87} &= AB \\
X_{53}=X_{57}=X_{61}=X_{65}=X_{73}=X_{77}=X_{81}=X_{85} &= AA \\
X_{54}=X_{58}=X_{62}=X_{66}=X_{74}=X_{78}=X_{82}=X_{86} &= AAB
\end{aligned}$$

(18)

Even if n changes, the slice will not change, so we can multiply by the same matrix in Eq. (14) (the matrix elements are given by Eq. (15)). Therefore, the result becomes the matrix in Eq. (14) with an exponent of 3N. If diagonalization is possible,

$$Z = \prod_i [\psi_i] \prod_{\{i,j\}} [\phi_{i,j}] Tr(U^{-1}XU)^{3N} \ . \quad (19)$$

For a node system of arbitrary N, as each node is added, if it is possible to represent the system as a repetition of the same operation (general crystal systems such as 3-dimensional lattices), we create $2^m$ versions of the matrix in Eq. (14), where m is the number of the slice (each element is created following the rules in Eqs. (16) and (17)), and diagonalize the matrix as follows to derive the partition function for the scale N→∞.

$$Z = \prod_i [\psi_i] \prod_{\{i,j\}} [\phi_{i,j}] Tr(U^{-1}XU)^N \ . \quad (20)$$

In the Ising model for general arbitrary shapes, because the number of connections for node n-1 is different from that of node n, the number of elements in the rows and columns of matrix X for node n are different. However, even in this case, we can create the matrix shown in Eq. (14) with different numbers of elements in the rows and columns, and by determining each element based on Eqs. (16) and (17), the partition function Z is determined by the product of the matrices. We derive the partition function for the following system as an example.

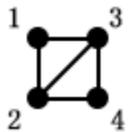
$(1+A_{12})(1+A_{13})(1+A_{23})(1+A_{24})(1+A_{34})$
$(1+B_1)(1+B_2)(1+B_3)(1+B_4)$

(21)

$$X_1 = \begin{bmatrix} 1 \\ A_{1,2}B_1 \\ A_{1,3}B_1 \\ A_{1,2}A_{1,3} \end{bmatrix}, X_2 = \begin{bmatrix} 1 & B_2 & A_{2,3}B_2 & A_{2,3} \\ A_{2,3}B_2 & A_{2,3} & 1 & B_2 \\ A_{2,4}B_2 & A_{2,4} & A_{2,3}A_{2,4} & A_{2,3}A_{2,4}B_2 \\ A_{2,3}A_{2,4} & A_{2,3}A_{2,4}B_2 & A_{2,4}B_2 & A_{2,4} \end{bmatrix},$$

$$X_3 = \begin{bmatrix} 1 & B_3 & A_{3,4}B_3 & A_{3,4} \\ A_{3,4}B_3 & A_{3,4} & 1 & B_3 \end{bmatrix}, X_4 = \begin{bmatrix} 1 & B_4 \end{bmatrix}. \qquad (22)$$

$$Z = \prod_i [\psi_i] \prod_{\{i,j\}} [\phi_{i,j}] (X_4 X_3 X_2 X_1) . \qquad (23)$$

As shown here, we assign numbers to each node, determine which other nodes interact with each node, and derive the partition function as the product of matrices that are created automatically. In statistical mechanics, the partition function for a system of an arbitrary shape in a magnetic field is derived using a method of solving Bayesian networks that contain loops, as described next.

*2.3 Bayesian network*

Calculation methods for Bayesian networks that contain loops are not yet known. However, it is possible to calculate them based on the method for solving the partition function of a general Ising model with a magnetic field.

$$P = \sum_{\{true,false\}} \cdots \sum_{\{true,false\}} \Pr\{1,2,\cdots\} . \qquad (24)$$

$$P = \sum_{\{true,false\}} \cdots \sum_{\{true,false\}} \prod_{\{i,j\}} \Pr\{i|j\} . \qquad (25)$$

{ij} is multiplied for all edges. Eq. (25) becomes the following: 2)

$$P = \frac{exp(\sum_{\{i|i\in V\}} h_i x_i + \sum_{\{i,j\}\in E} J_{i,j} x_i x_j)}{\sum_{\vec{z}_V} exp(\sum_{\{i|i\in V\}} h_i x_i + \sum_{\{i,j\}\in E} J_{i,j} x_i x_j)} . \qquad (26)$$

$$h_i = \frac{1}{4} \sum_{\{j|\{i,j\}\ni i\}} \sum_{z_i=\pm 1} \sum_{z_j=\pm 1} z_i \ln f_{\{i,j\}}(z_i, z_j) . \qquad (27)$$

$$J_{i,j} = \frac{1}{4} \sum_{z_i=\pm 1} \sum_{z_j=\pm 1} z_i z_j \ln f_{\{i,j\}}(z_i, z_j) . \qquad (28)$$

This is the partition function for an Ising model of an arbitrary shape in a magnetic field. Therefore, we obtain a solution method for arbitrary Bayesian networks containing loops, based on the solution method described earlier. By obtaining a method for solving Ising models in magnetic fields, we also obtain a method for solving Bayesian networks, which is undoubtedly relevant in the field of solid-state physics. This method is anticipated to have applications in the fields of data mining, machine learning, and combinatorial optimization, which are based on Bayesian networks. In addition, because this method makes it possible to obtain the partition function for arbitrary crystal systems, we also aim to pursue applications of this method in the field of complex systems, such as phase transitions and emergence, in the future.

## 3. Results

The partition function of an arbitrary shape (structure of nodes and edges) in a magnetic field is given by the product of the matrices that are created according to the rules shown in Eqs. (14), (15), (16), and (17). If the system repeats some regular pattern, like a crystal, then it can be represented as the repetition of the product of the same matrices. The matrices can be diagonalized and represented as a matrix raised to the power of N. Then, the partition function for the scale $N\to\infty$ can be derived. In addition, if we want to consider the case in which there is no magnetic field, then we only need to set $B = 0$ in the results. In addition, the method for solving the partition function of an arbitrary shape in a magnetic field can be used to calculate Bayesian networks with loops. Regardless of the type of network, it is possible to analyze the network using a Bayesian network by tracing each individual node and determining which nodes it is connected to.

E-mail: saito@tsunagari-llc.jp